\documentclass[aps,pre,reprint,groupedaddress]{revtex4-2}

\usepackage{color}
\usepackage{amsmath}
\usepackage[pdftex]{graphicx}
\usepackage{url}

\begin{document}


\title{Graph-based open-ended survey on concerns related to COVID-19}


\author{Tatsuro Kawamoto}
\affiliation{  Artificial Intelligence Research Center, \\
  National Institute of Advanced Industrial Science and Technology, 
  Tokyo, Japan }

\author{Takaaki Aoki}
  \affiliation{Faculty of Education, Kagawa University, Takamatsu, Japan}


\author{Michiko Ueda}
\affiliation{Faculty of Political Science and Economics, Waseda University, Tokyo, Japan }



\begin{abstract}
The COVID-19 pandemic is an unprecedented public health crisis with broad social and economic consequences. We conducted four surveys between April and August 2020 using the graph-based open-ended survey (GOS) framework, and investigated the most pressing concerns and issues for the general public in Japan. The GOS framework is a hybrid of the two traditional survey frameworks that allows respondents to post their opinions in a free-format style, which can subsequently serve as one of the choice items for other respondents, just as in a multiple-choice survey. As a result, this framework generates an opinion graph that relates opinions and respondents. We can also construct annotated opinion graphs to achieve a higher resolution. By clustering the annotated opinion graphs, we revealed the characteristic evolution of the response patterns as well as the interconnectedness and multi-faceted nature of opinions. Substantively, our notable finding is that ``social pressure,'' not ``infection risk,'' was one of the major concerns of our respondents. Social pressure refers to criticism and discrimination that they anticipate receiving from others should they contract COVID-19. It is possible that the collectivist nature of Japanese culture coupled with the government's policy of relying on personal responsibility to combat COVID-19 explains some of the above findings, as the latter has led to the emergence of vigilantes. The presence of mutual surveillance can contribute to growing skepticism toward others as well as fear of ostracism, which may have negative consequences at both the societal and individual levels.
\end{abstract}


\maketitle

\section*{Introduction}
The COVID-19 pandemic is an unprecedented event with a myriad of consequences. Without a doubt, it is one of the most serious public health crises in recent history. However, it is more than a public health challenge. Various measures that were introduced to prevent the spread of the virus have caused major disruptions in economic activities and social life. Reduced business activities and associated job losses, school closures, movement restrictions, and social distancing have affected multiple aspects of our lives, presenting new challenges that every single member of society has to face in addition to the disease itself. Thus, the COVID-19 pandemic has given rise to multi-faceted issues and concerns for many individuals, ranging from the consequences of the infection itself to the economic and social ramifications of the pandemic. 
Throughout this paper, we refer to these concerns and issues as opinions. 

Using the graph-based open-ended survey (GOS) framework \cite{KawamotoAoki2019}, this paper seeks to understand such multi-faceted opinions expressed by the general population in Japan between April and August 2020. The GOS framework is a hybrid of the two traditional survey frameworks: multiple choice and free format. In a multiple-choice survey, respondents are presented with a list of items from which they could choose one or more that were most applicable to them. While this is a widely used framework owing to its ease of implementation, the potential answers must be anticipated and compiled as a list by those conducting the survey. Thus, the multiple-choice survey may not be the most appropriate framework to use if the purpose is to gauge complex and multi-faceted opinions, such as people's concerns during an unprecedented event like the COVID-19 pandemic. Their concerns and opinions may not be clear a priori because of the complex and rapidly developing nature of the event. The second framework is the free-format survey in which the respondents post their opinions as texts. However, analysing such texts can be arbitrary to a certain extent and challenging in a large-scale survey, including this study. 

In the GOS framework, respondents can post their opinions in a free-format style, which can subsequently serve as one of the choice items for other respondents, just as in the multiple-choice survey. Because survey respondents can select others' opinions that they agree with, the GOS framework constitutes a graph (or network) of responses at the end. This graph makes statistical evaluation of the free-format responses feasible, while preserving the diversity of respondents' opinions that increases spontaneously as respondents express their own opinions. 

\begin{figure}[ht!]
  \begin{center}
    \includegraphics[width=0.9\linewidth]{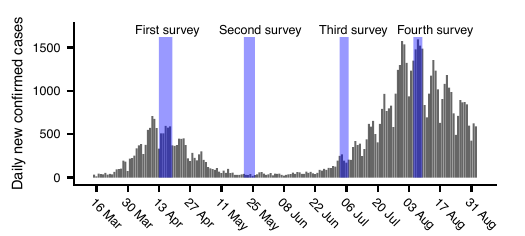}
    \caption{Survey periods and the daily new confirmed cases of COVID-19 in Japan. }
    \label{fig:daily_cases}
  \end{center}
\end{figure}

A major contribution of this paper is to reveal the most pressing issues and concerns of the general public through the opinions expressed by the respondents. In addition, by conducting the same survey using the GOS framework several times, we were able to capture the evolution of those opinions. 

Our analysis shows that the variety of major opinions rapidly changed during the study period, and that several opinions were strongly related. Traditional survey methods are less suitable to capture such characteristic behaviours.
These results indicate that studying opinions under a rapidly changing situation, such as the COVID-19 pandemic, is indeed an ideal application of the GOS framework.

Our study period corresponded to the middle of the first wave of cases in Japan (April 2020), the period in which the number of cases temporarily diminished (May 2020), and the beginning and the middle of the second wave (July and August 2020, respectively), which was larger than the first one in terms of the number of cases, but resulted in fewer deaths (see Fig.~\ref{fig:daily_cases} for the evolution of the confirmed cases). The first case of COVID-19 in Japan was reported on January 16. The Japanese government declared a state of emergency in major metropolitan areas on April 7, and expanded the coverage to the entire country on April 16. The Japanese government implemented neither a lockdown nor strict movement restrictions. Non-essential businesses were urged to either temporarily close or operate for reduced hours, but non-compliance was not penalised. The state of emergency was lifted on May 25. As of August 31, the total number of positive cases was 67,077, and the number of COVID-19-related deaths was 1,278, which amounted to 10.15 deaths per one million people \cite{ministry_of_health_labour_and_welfare_number_2020}.

\section*{Materials and methods}
We conducted four rounds of an online survey of the Japanese adult population aged 20--59 between April and August 2020. The respondents were asked about their most pressing concerns. Specifically, the question asked (in Japanese) was, ``What is the concern and the difficulty that you face in your daily life and economic activities? Please provide concrete and succinct answers.'' The first survey was conducted between April 13 and 19, 2020. The subsequent three surveys were conducted on the following dates: May 21--26, July 3--7, and August 5--9, 2020. 
In each round, a set of screening questions was sent to approximately 10,000 individuals who are members of commercial web panels.
We then selected a sample of approximately 6,000 respondents each time, based on their demographic characteristics, to represent the Japanese general population in terms of their residency area, sex, and age groups. An invitation to our survey was sent to these selected individuals through a survey company. We used all the responses in our analysis and no further criteria has been applied to screen the respondents. The number of respondents who participated in the four rounds of the survey were 2103, 1516, 1729, and 1659, respectively.  Their demographic attributes are reported in Table \ref{tab:Demographic} in Supplementary Information. 

The survey participants were informed of the purpose of the study prior to their participation, and had the option to quit at any time. The respondents provided explicit digital consent that the information they provided could be used for the purpose of this study. The data were completely anonymous. This study, including the use of digital consent, was approved by the Ethics Review Committee on Human Research of Waseda University (approval number: 2020-050). 

\subsection*{Graph-based open-ended survey}\label{sec:GOSsampling}
The results of the GOS were used to generate a bipartite graph consisting of two sets of vertices representing the opinions and respondents, respectively. 
Each edge connects a pair of opinions and respondent vertices, and indicates that a respondent supports the opinion. 
We refer to this graph as an \textit{opinion graph} (see Supplementary Information for a comment on the definition of the opinion graph).

\begin{figure*}[t!]
  \begin{center}
    \includegraphics[width=0.8\linewidth]{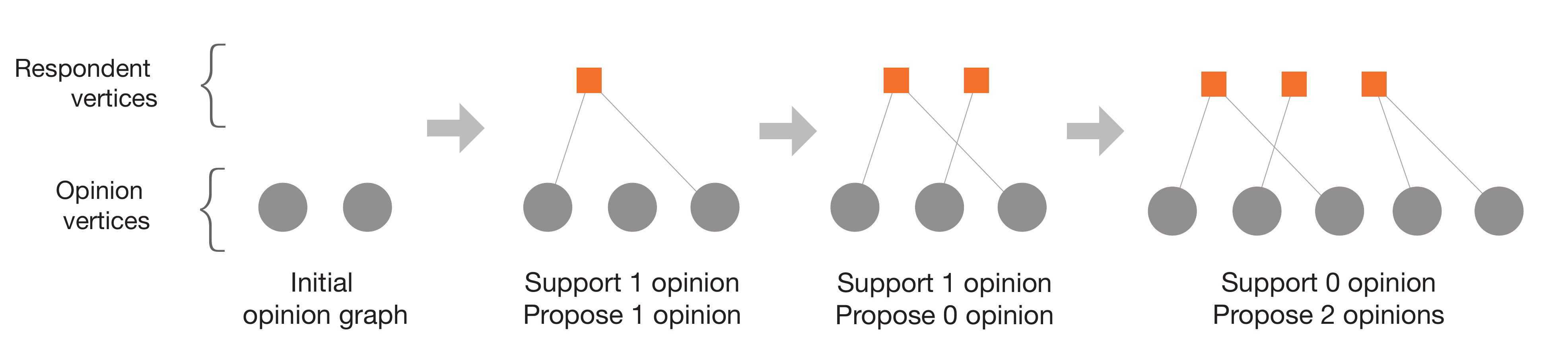}
    \caption{
    Example of the generation process of an opinion graph.
    }
    \label{fig:Schematic}
  \end{center}
\end{figure*}

The generation process of the opinion graph is illustrated in Fig.~\ref{fig:Schematic}.
At the beginning of the survey, we prepared a set of opinion vertices (which can be an empty set) as the initial opinions (choices). The description of the initial opinions is provided in Supplementary Information. When a respondent responds to a question, several opinions are sampled uniformly and randomly, and presented to the respondent; the minimum number of samples is eight, while respondents can opt to refer to up to 24 opinions. 
Whenever a respondent supports a presented opinion(s), the GOS generates a respondent vertex connected to the selected opinion vertices. 
When a respondent expresses new opinions, the GOS generates corresponding opinion vertices connected to the newly added respondent vertex. 
The numbers of opinion vertices and respondent vertices increase simultaneously as the GOS is carried out. The above procedure constitutes an opinion graph. 

We expect a set of similar opinion vertices to be connected via respondent vertices. 
Similarly, we expect the respondent vertices to be classified into groups, in which the vertices have a similar response pattern. 
The classification of opinions and respondents is realised through clustering of an opinion graph, which we refer to as the opinion groups and respondent groups, respectively.

\subsection*{Clustering of the opinion graphs}\label{sec:GraphClustering}
A number of clustering algorithms have been proposed for graphs and bipartite graphs in the literature \cite{Dhillon2001,NMF,Kluger2003,Larremore2014,Gerlach2018}. 
Here, we used the Markov chain Monte Carlo method, which was implemented in a software called \textit{graph-tool} \cite{graphtool} that performs a Bayesian inference of groups under the assumption of a stochastic block model \cite{Peixoto2017tutorial}. 
The algorithm identifies a statistically significant group structure and estimates the number of groups in a non-parametric manner. 
All the clustering results in this study are inferences of unnested (i.e. non-hierarchical) stochastic block models. 
We confirmed that, while the nested variant does not raise the resolution considerably, it often subdivides a large opinion group into smaller sets, which is apparently an overfit.

\subsection*{Annotated opinion graphs}\label{sec:AnnotatedGraphs}
In addition to the clustering of raw opinion graphs, we also considered annotated opinion graphs to achieve a higher resolution by adding more information externally through annotation. 

Based on the collected opinions, we asked three annotators to classify the opinions into 10 groups: 
(1) \textit{infection risk}, 
(2) \textit{social pressure \& future prospect}, 
(3) \textit{financial issues}, 
(4) (\textit{restriction of}) \textit{travel}, 
(5) \textit{government policies}, 
(6) \textit{mask} (\textit{shortage}), 
(7) \textit{mask} (\textit{discomfort}), 
(8) \textit{other issues}, 
(9) \textit{no concerns}, and 
(10) \textit{invalid responses}.
The opinions that may not exclusively belong to one of these groups are left unannotated. The last item refers to opinions that are not directly related to the question that was asked. More detailed descriptions of these groups are provided in Supplementary Information.  

The annotators classified opinions independently, and we distinguished annotations according to different annotators. The maximum number of annotation labels was 30 for each survey. We denoted the total number of annotation labels as $K$, which can be less than 30 if some labels are not used.

\begin{figure*}[t!]
  \begin{center}
    \includegraphics[width= 0.99\linewidth]{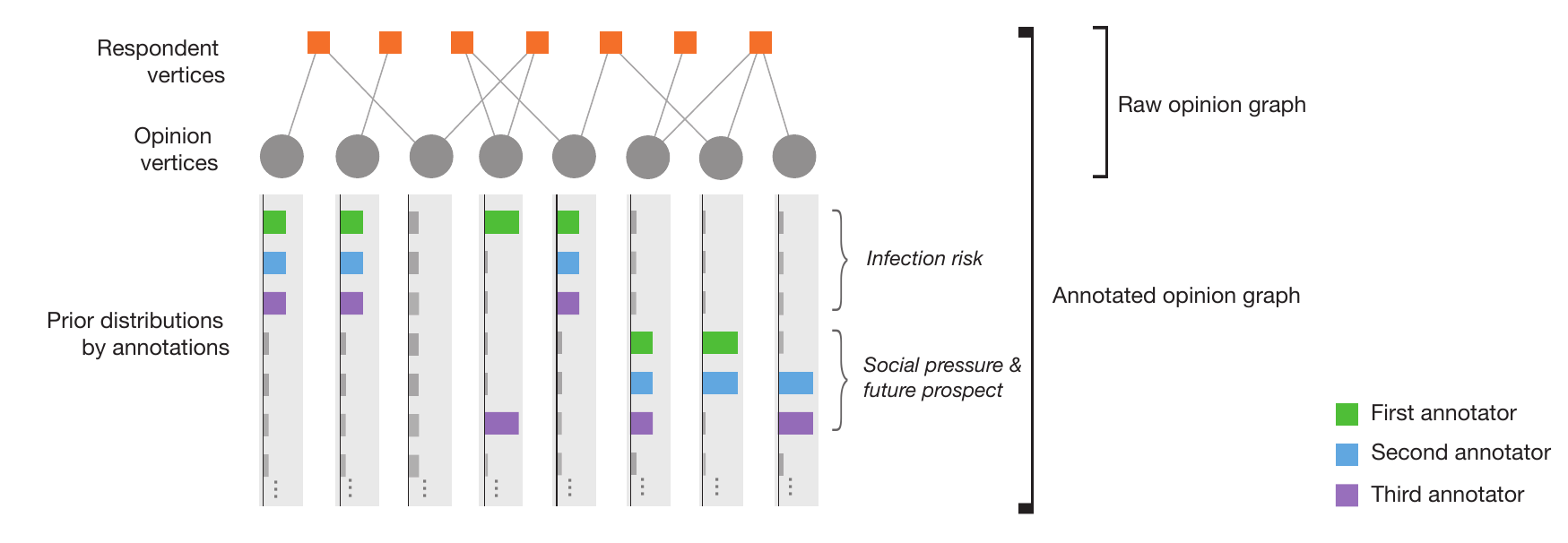}
    \caption{Schematic of an annotated opinion graph. A raw opinion graph is shown at the top, and the prior distribution for each vertex is shown as a bar plot at the bottom. }
    \label{fig:AnnotationSchematic}
  \end{center}
\end{figure*}

As shown in Fig.~\ref{fig:AnnotationSchematic}, based on annotators' decisions, we constructed a $K$-dimensional vector that represents a prior probability distribution of the group assignment for each opinion vertex. The $k$th element is $\eta$ ($\gg 0$) if the vertex has the $k$th annotation label; otherwise, $\epsilon$ ($0< \epsilon \ll \eta$), where $\epsilon$ and $\eta$ are determined to normalise the vector to unity. 
Therefore, the annotated vertices have biased prior distributions, whereas all unannotated vertices have a uniform prior distribution.

These prior distributions were incorporated into the Bayesian inference using the \textit{bfield} parameter in \textit{graph-tool}. 
The set of opinion vertices with similar prior distributions tends to be classified into the same group. 
Because we used a non-parametric inference method, the algorithm eventually identifies the parsimonious number of opinion groups.
Alternatively, one could use an approach discussed in \cite{Hric_PRX2016} to incorporate annotations in the inference.

\section*{Results}
\subsection*{Response patterns}\label{sec:ResultsPopularityMatrices}

In each round, 191 (9.1\%), 127 (8.4\%), 119 (6.9\%), and 117 (7.1\%) respondents expressed their own opinions, while the remaining respondents selected opinion(s) that were applicable to them from a list presented to them. 

\begin{figure*}[t!]
  \begin{center}
    \includegraphics[width=\linewidth]{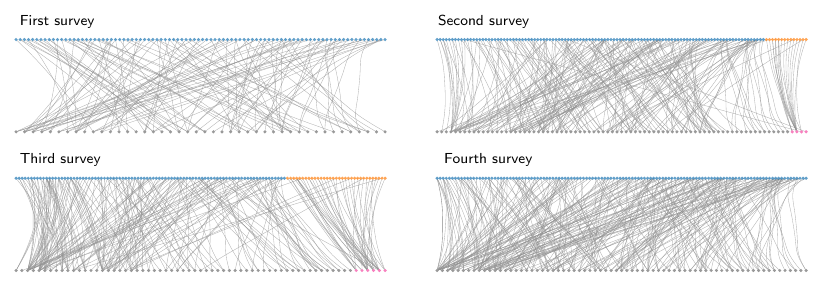}
    \caption{
    \textbf{Clustering results of the opinion graphs.} 
    These graphs only show randomly sampled connected subgraphs instead of the entire dataset, for illustration purposes. 
    The opinion vertices are aligned at the bottom, while the respondent vertices are aligned at the top. Both types of vertices are sorted based on the inferred group assignments. 
    The colour of each vertex represents the group assignment. 
    The pink vertices are the set of opinions that can be coded as \textit{no concerns}.
    }
    \label{fig:RawOpinionGraphs}
  \end{center}
\end{figure*}

Fig.~\ref{fig:RawOpinionGraphs} shows the clustering results of the raw opinion graphs. 
For the first and fourth surveys, no group structure was identified. 
In contrast, for the second and third surveys, opinion graphs were partitioned into two groups. In both cases, the vertices in the smaller respondent group are highly connected to the opinion vertices that can be coded as \textit{no concerns}, while the respondents in the other group supported the opinions on various issues, such as infection risk and financial concerns.

\begin{figure*}[ht!]
  \begin{center}
     \includegraphics[width=0.9\linewidth]{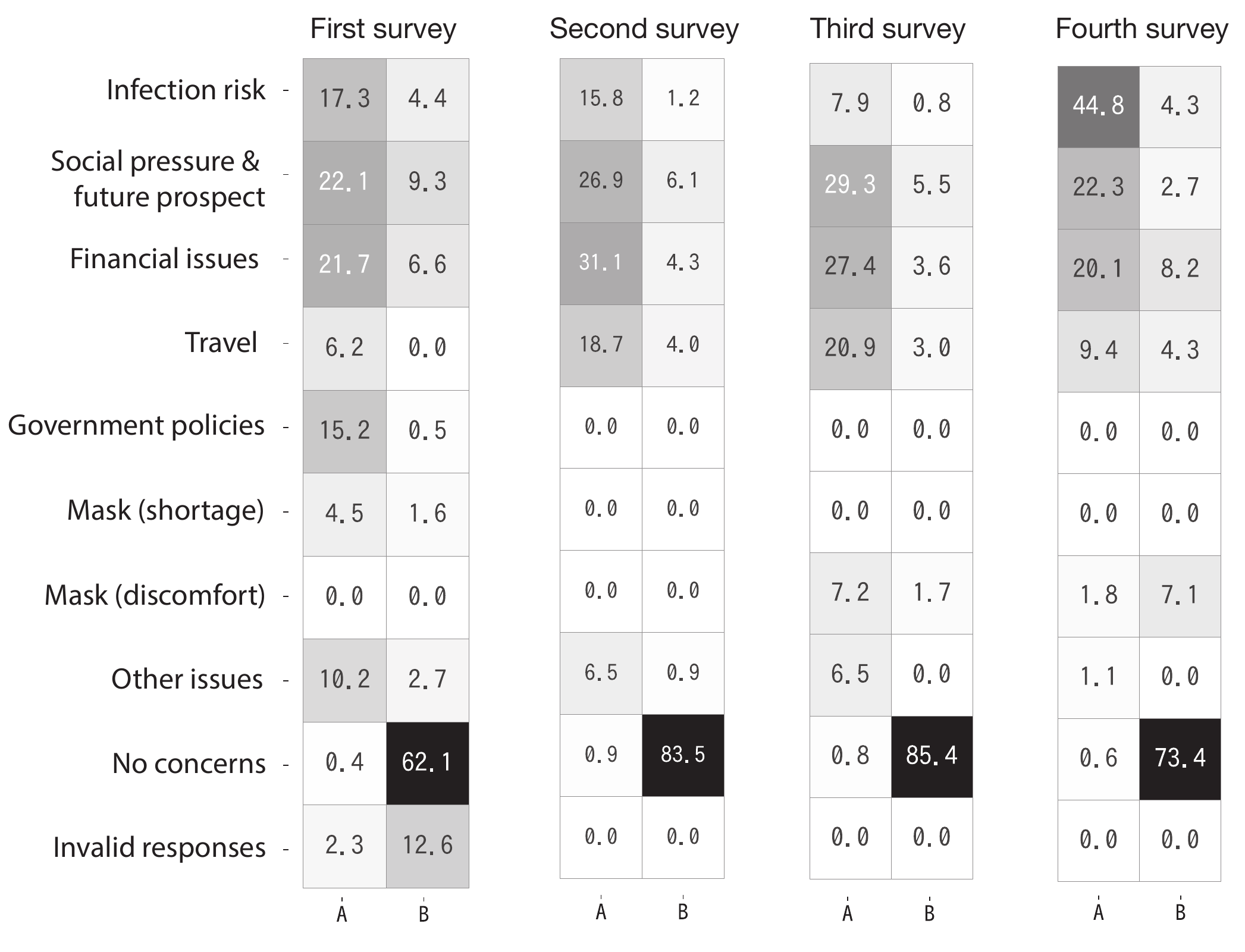}
    \caption{
    Response patterns of the four surveys. 
    The rows indicate opinion groups and the columns indicate respondent groups. In each table, the two respondent groups are indicated as groups A and B.
}
    \label{fig:PopularityMatrix}
  \end{center}
\end{figure*}

The clustering in Fig.~\ref{fig:RawOpinionGraphs} is the most agnostic result based purely on the collected responses. However, their resolution is insufficient, as the raw opinion graph is only classified into a few groups or not classified at all. 
Fig.~\ref{fig:PopularityMatrix} shows the response patterns of respondents that were obtained through the clustering of annotated opinion graphs. 
The $(i,j)$ element is the group-wise propensity (normalised in each column) towards opinion group $i$ for respondent group $j$. More precisely, this is the fraction of the number of edges between groups $i$ and $j$.
Although the number of identified opinion groups in each survey was nine (first survey), six (second survey), and seven (third and fourth surveys), we added empty opinion groups in Fig.~\ref{fig:PopularityMatrix} to obtain the same number of rows in all surveys. 
In contrast to the results reported in Fig.~\ref{fig:RawOpinionGraphs}, where we observed only one group of respondents for the first and fourth surveys, we can now observe two groups analogous to the results of the second and third surveys: respondents who had specific concerns (group A) and those who did not (group B).

\begin{figure*}[ht!]
  \begin{center}
     \includegraphics[width=0.75\linewidth]{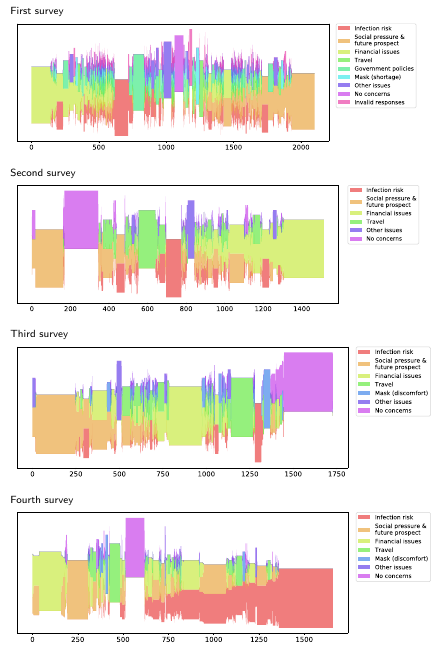}
    \caption{
    Palette diagrams of the four surveys. 
}
    \label{fig:PaletteDiagrams}
  \end{center}
\end{figure*}

To further reveal the details about response patterns, we show the palette diagrams \cite{Noguchi2019,Noguchi2020} in Fig.~\ref{fig:PaletteDiagrams}. 
The palette diagram is essentially a \textit{streamplot}, which is a stack plot with varying origin axes. 
The streamplot is a common visualization for time-series data, for example, the evolution of car companies' market share. In the palette diagram, the vertical axis represents the normalized propensity (i.e., vertical thickness represents unity) for each opinion group that a respondent supports. 
The set of normalised propensity patterns for all respondents are optimally aligned in a horizontal fashion, so the global distribution of the response pattern is better understood. 
In other words, the respondent vertex indices correspond to the timestamps in the time-series data. The palette diagrams in Fig.~\ref{fig:PaletteDiagrams} show the microscopic response patterns of respondents. The tables in Fig.~\ref{fig:PopularityMatrix} show how these patterns can be summarised at a macroscopic level.

The results presented in these two figures suggest that there was a sizable block of respondents who did not express any specific concerns throughout the study period. These respondents were identified as a distinct group, relatively isolated from others, through graph clustering. The evolution of the number of respondents in this group is negatively correlated with the number of daily cases, which peaked during our fourth survey, as expected. As indicated in Fig.~\ref{fig:PaletteDiagrams}, these respondents were less likely to support other opinion groups. 

For the other opinion groups, it should be noted that many of these groups, including \textit{government policies} and \textit{travel}, emerged spontaneously from the opinions expressed by respondents themselves, as they were not included in the initial opinions displayed at the beginning of the survey. On the other hand, the opinions belonging to \textit{ infection risk}, \textit{ social pressure \& future prospect}, and \textit{ financial issues} were included in the initial opinions. It should also be noted, however, that the initial opinions become less likely to be displayed to the respondents as more respondents post their own opinions. Therefore, the initial opinions are unlikely to have a significant influence on the result. 

In all four surveys, the opinions of respondent group A (shown in Fig.~\ref{fig:PopularityMatrix}) tended to focus on the following opinion groups: \textit{infection risk}, \textit{social pressure \& future prospect}, \textit{financial issues}, and \textit{travel}. The size of these opinion groups varied considerably across surveys. For example, \textit{travel}, which covers issues associated with difficulties with traveling and stay-at-home ``requests,'' was not a major issue in April, as there were no formal movement restrictions in Japan. It became more prevalent in the second and third surveys (May and July) when the number of cases diminished temporarily. By August, however, it no longer constituted a major opinion as the infection risk again became the dominant concern with the rising number of cases. 

In the first survey (April 2020), our respondents also tended to complain about the government's handling of the pandemic (\textit{government policies}) as well as express concerns regarding the status of hospitals that were overwhelmed by the rapidly rising number of patients; this was captured in the \textit{other issues} group. However, as indicated in Fig.~\ref{fig:PopularityMatrix}, the fraction of such concerns quickly decreased in subsequent surveys as the focus of respondents shifted to other issues, such as financial concerns. 

It is noteworthy that \textit{social pressure \& future prospect} always constituted one of the major opinion groups throughout our study period. This group mainly refers to the negative \textit{social} consequences of contracting the disease, rather than the physical consequences. Many respondents expressed fears that they might be ostracised by others if they contract the disease; thus, this opinion group largely represents the social pressure that respondents felt even before testing positive for COVID-19. Similarly, \textit{financial issues} was another group that appeared as a sizable block in all of the surveys, indicating that the economic consequences of the pandemic were among the major concerns of respondents. 

Fig.~\ref{fig:PaletteDiagrams} reveals that many respondents expressed multiple concerns and opinions, instead of selecting a single issue as their most pressing concern. While \textit{financial issues}, \textit{social pressure \& future prospect}, and \textit{no concerns} tended to appear on their own, other types of issues were more likely to appear in combination with other opinion groups, suggesting that many of our respondents faced multiple issues and concerns during the pandemic. 

Finally, our results show that infection risk became the dominant concern by the time we conducted the fourth survey, which coincided with the peak period of the second wave of cases (August). 
As shown in Fig.~\ref{fig:PaletteDiagrams}, although the \textit{infection risk} group was one of many other opinions for most respondents until the third survey, it became a prominent group in the fourth survey. The number of respondents who specified infection risks as their only concern was highest in the fourth survey, which suggests that infection risk became the dominant issue for a large number of respondents.

\section*{Conclusion and Discussion}
In this paper, we surveyed the general adult population of Japan four times using the GOS framework to investigate the issues and concerns of people in the midst of the COVID-19 pandemic. 
We classified the opinions of respondents by aggregating the response patterns and annotations through graph clustering. As a result, we revealed the characteristic evolution of the response patterns, particularly on infection risk.

The survey results indicated that the most pressing concerns of the general public changed through the various phases of the pandemic. Many of our respondents also had multiple issues and concerns. Such a fine evolution of people's opinions as well as the multi-faceted nature of these opinions may not be adequately captured by the traditional multiple-choice survey method, which required us to predict the opinions to be included in a list of potential choices. Presenting survey respondents with a pre-determined set of responses can be particularly challenging when the situation changes on a daily basis, as in the case of the current pandemic. Therefore, this is an ideal situation for an open-ended survey, in which the group labels are determined a posteriori. 
As a hybrid of the two traditional survey methods, the GOS framework takes advantage of both methods and successfully reveals the multi-faceted and rapidly changing features of people's opinions. 

Substantively, our notable finding is that ``social pressure,'' not ``infection risk,'' was one of the major concerns of our respondents. Social pressure refers to criticism and discrimination that they anticipate receiving from others should they contract COVID-19. Although stigmatization and social rejection of those with an infectious disease have been reported both in past pandemics and the current one \cite{siu_sars-associated_2008, lee_experience_2005, bagcchi_stigma_2020}, it is noteworthy that individuals in the general population, who were neither actual COVID patients nor high-risk people (e.g., healthcare workers), widely anticipated and worried about stigmatization and ostracism even when their chance of contracting COVID was infinitesimally small, as in our study period. 

While a full explanation of the factors underlying such anticipated social stigma is beyond the scope of the current paper, it is possible that the collectivist nature of Japanese culture coupled with the government's policy of relying on personal responsibility to combat COVID-19 explains some of the above findings, as the latter has led to the emergence of vigilantes. In collectivism, people tend to place greater value on group objectives than on personal ones \cite{markus_culture_1991}; thus, those who do not prioritize group goals (in this particular context, public health) are more likely to be ostracized and excluded. In addition, the Japanese government's emphasis on self-restraint, instead of introducing strict official rules, created an environment in which vigilantes, or self-appointed police, emerged and were actively engaged in reporting and attacking people not complying with the government ``guidelines.''

The activities of vigilantes, for whom the special term ``\textit{jishuku keisatsu}'' (self-restraint police) was coined, were widely reported during the pandemic in Japan. They included instances of harsh criticism, harassment, and vandalism toward those who were considered ``rule-breakers.'' For instance, vehicles from non-local areas were vandalized when travel restraint requests were in place; stores that did not follow the government's request to operate at reduced hours were anonymously threatened, and COVID-19 patients who did not comply with the government guidelines were severely criticized \cite{tomohiro_osaki_japans_2020, patrick_michel_understanding_2020}.

In addition, the Japanese tend to blame COVID patients for contracting the disease, and one study reported that people in Japan were much more likely to think that it was a COVID patient's fault that they contracted the disease, citing their own responsibility, compared to those in other developed countries  \cite{miura_2020}. Thus, our respondents might also have expected that they would be blamed for their action, because people would be likely to perceive it as a consequence of their action and non-compliance.

Such an attitude and the presence of mutual surveillance can contribute to growing skepticism toward others as well as fear of ostracism, which may have negative consequences at both the societal and individual levels. This is especially true as rejection sensitivity is known to be relatively higher in Japan \cite{garris_consequences_2011}. There was even one instance of suicide by a COVID patient in January 2021, who was supposedly motivated by self-blame and the fear of ostracization by others \cite{rich_pandemic_2021}.

Our findings suggest that many members of Japanese society felt pressure associated with COVID-related social rejection during the early phase of the pandemic. Stigmatization and blaming of non-compliers may have detrimental effects on the general population, especially in a collectivist culture, but we have a very limited understanding of their effects, let alone their existence, during the current pandemic. Thus, highlighting the significance of social pressure for members of a collectivist society during a public health crisis is an important substantive contribution of the present study. Understanding the potential psychological effects of social pressure and conducting cross-cultural comparisons of such effects constitute an important future research agenda.

Finally, let us discuss the role of annotation in the clustering of opinion graphs. 
If every respondent supports only one issue or concern, then the resulting opinion graph would have a relatively simple group structure, in which each respondent group would be densely connected to the corresponding opinion group. However, as observed in Figs~\ref{fig:PopularityMatrix} and \ref{fig:PaletteDiagrams}, the opinion graphs in the present survey have more complex group structures because a significant number of respondents support multiple opinions with different meanings. 
The annotations were needed to achieve a higher resolution because a graph with a complex group structure is difficult to distinguish from a uniformly random graph, that is, pure noise.

Although the annotations made the group structures clear, it should also be noted that clustering annotated opinion graphs is far from trivial. 
First, because the annotators' classifications were performed independently, just as the responses of the surveys, the decisions of the annotators are subjective and do not coincide precisely (see Supplementary Information for the consistency among the annotators). 
In addition, not all opinion vertices were annotated, and there were no annotations for respondent vertices. 
Considering the fact that the responses to open-ended questions are often classified (i.e. annotated) in a completely manual manner, our statistical inference approach that brings together all the information from respondents and annotators is a considerable improvement.

\begin{acknowledgments}
This study was financially supported by JSPS Grants-in-Aid for Scientific Research Grant Nos. 18K18604 (T.A. and T.K.) and 20H01584 (M.U.), and Research and Regional Cooperation for Crisis Management Shikoku (T.A.). The funders had no role in study design, data collection and analysis, decision to publish, or preparation of the manuscript.
\end{acknowledgments}


\clearpage

\pagebreak
\begin{center}
\textbf{\large Supplementary Information: \\Graph-based open-ended survey on concerns related to COVID-19}
\end{center}
\setcounter{equation}{0}
\setcounter{figure}{0}
\setcounter{table}{0}
\setcounter{section}{0}
\setcounter{page}{1}
\makeatletter
\renewcommand{\thesection}{S\arabic{section}}
\renewcommand{\theequation}{S\arabic{equation}}
\renewcommand{\thefigure}{S\arabic{figure}}
\renewcommand{\bibnumfmt}[1]{[S#1]}
\renewcommand{\citenumfont}[1]{S#1}

\section*{On the definition of the opinion graph}\label{OpinionGraphDefinition}
The data collection procedure in this study is essentially the same as the survey conducted in \cite{KawamotoAoki2019}. However, we considered bipartite opinion graphs, and treated both opinion and respondent vertices as vertices of the same type (i.e. monopartite graph) in \cite{KawamotoAoki2019}. 
Note that we only considered the single-edge type in this study for simplicity, whereas we considered multiple-edge types (positive and negative) in \cite{KawamotoAoki2019} for graph clustering.

\section*{Initial opinions}\label{InitialOpinion}

Below is a list of the initial opinions that were randomly presented to respondents. The original texts were written in Japanese. 

\begin{itemize}

\item Domestic violence
\item I had to close my business because of a declining number of customers.
\item I cannot concentrate on my work because of school closure.
\item I am afraid of the pressure that I might experience from others if I become infected with COVID-19.
\item My business is suffering. 
\item I do not know what is going to happen to me if I become infected. 
\item I am hesitant to visit my doctor even when I feel sick with the common cold, etc., because I do not want to catch the virus.
\item I cannot get a PCR test even when I want to get tested. 
\item I cannot get a paid leave even if I tested positive with COVID-19, and I would have no choice but to take a leave without pay or quit my job. 
\item I am not sure if I can afford to pay medical bills if I become infected. 
\item I am spending too much on childcare costs.
\item I do not have much money left and I am not sure if I can survive without public assistance. 
\end{itemize}

These opinions were selected because they could be considered typical opinions. We also included several opinions from each of the potential major categories: infection risk, social pressure/future prospect, and financial issues. 

We provided these initial opinions to help respondents understand the scope of the question. As pointed out in \cite{Schuman:1987cv}, providing the scope of a question is important for an open-ended question, and it is often mentioned in the question. If no information about the scope was provided, then the diversity of opinions could be narrower than what could have been obtained. 

A drawback of the initial opinions is that they may influence respondents' opinions. However, the contents of the resulting opinion groups sufficiently deviated from the initial opinions; thus, it is unlikely that these initial opinions influenced the results of the present study.

\section*{Opinion groups}\label{sec:OpinionGroups}
Brief descriptions of each opinion group are provided below.
\begin{description}
\item[infection risk] 
Concerns related to the infection risk of COVID-19
\item[social pressure \& future prospect] Fears that they may be criticised should they be tested positive for COVID-19; concerns about uncertain future prospects
\item[financial issues] Concerns about income and/or employment status
\item[travel] Dissatisfaction with difficulties in traveling or going out
\item[government policies] Issues related to one-time cash handouts, dissatisfaction with politicians
\item[mask (shortage)] Shortage of masks at pharmacies
\item[mask (discomfort)] Discomfort associated with wearing a mask
\item[other issues] This group contains various valid opinions that are not classified into other opinion groups identified in each survey. 
For example, we identified the following contents in this opinion group:
\begin{enumerate}
    \item Worries about the capacity of hospitals
    \item Frustration with the behaviour of others, or school
    \item Shortage of items in stores (other than masks); physical and mental health
    \item Family-related issues (e.g. parenting)
\end{enumerate}
\item[no concerns] No particular concerns about the present situation. 
\item[invalid responses] Responses that are not directly related to the question asked in the present survey
\end{description}

\section*{Consistency among the annotators}\label{sec:AnnotatorConsistency}
\begin{figure*}[ht!]
  \begin{center}
    \includegraphics[width= \linewidth]{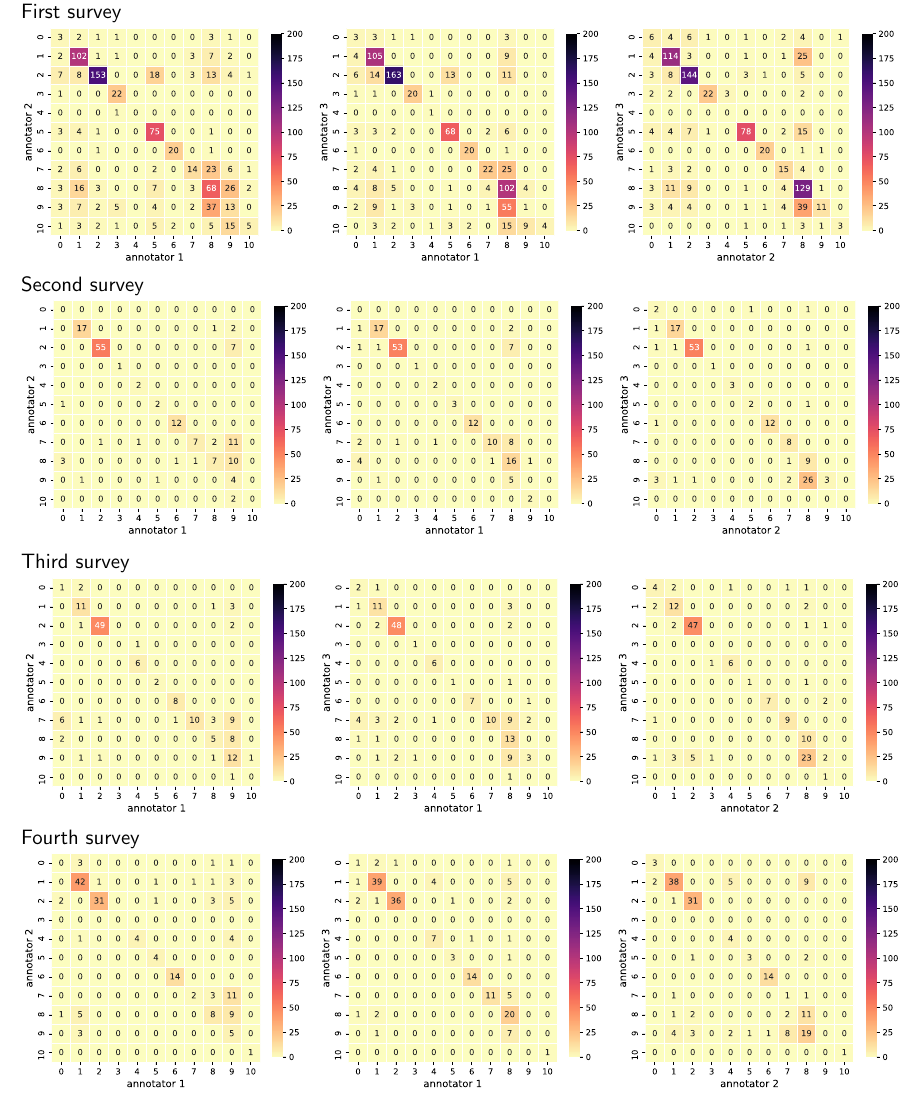}
    \caption{Agreement between each pair of annotators with respect to the opinion group labels.}
    \label{fig:AgreementMatrices}
  \end{center}
\end{figure*}

The consistency among the three annotators is shown in the tables in Fig.~\ref{fig:AgreementMatrices}.

\clearpage

\begin{table}
    \centering
    
    \begin{tabular}{lrrrrrrrrrr}
    \hline
         & \multicolumn{2}{c}{All} &  
           \multicolumn{2}{c}{1st survey} &  
           \multicolumn{2}{c}{2nd survey} &   
           \multicolumn{2}{c}{3rd survey} & 
           \multicolumn{2}{c}{4th survey} \\
         & N & \% & N & \% & N & \% & N & \% & N & \% \\ \hline
        \textbf{All} & 7007 & 100.0 & 2103 & 100.0 & 1516 & 100.0 & 1729 & 100.0 & 1659 & 100.0 \\
         &  &  &  &  &  &  &  &  &  &  \\ 
        \textbf{Sex} &  &  &  &  &  &  &  &  &  &  \\ 
        Male & 3928 & 56.1 & 1156 & 55.0 & 849 & 56.0 & 971 & 56.2 & 952 & 57.4 \\ 
        Female & 3079 & 44.0 & 947 & 45.0 & 667 & 44.0 & 758 & 43.8 & 707 & 42.6 \\ 
         &  &  &  &  &  &  &  &  &  &  \\ 
        \textbf{Age Groups} &  &  &  &  &  &  &  &  &  &  \\ 
        20-29 & 1028 & 14.6 & 429 & 20.4 & 231 & 15.2 & 195 & 11.3 & 173 & 10.4 \\ 
        30-39 & 1460 & 20.8 & 498 & 23.7 & 301 & 19.9 & 349 & 20.2 & 312 & 18.8 \\ 
        40-49 & 2268 & 32.4 & 572 & 27.2 & 448 & 29.6 & 610 & 35.3 & 638 & 38.5 \\ 
        50-59 & 2251 & 32.1 & 604 & 28.7 & 536 & 35.4 & 575 & 33.3 & 536 & 32.3 \\ 
         &  &  &  &  &  &  &  &  &  &  \\ 
        \textbf{Occupation/Employment Status} &  &  &  &  &  &  &  &  &  &  \\ 
        Working full-time & 3818 & 54.5 & 1128 & 53.6 & 808 & 53.3 & 946 & 54.7 & 936 & 56.4 \\ 
        Housekeeping and working & 552 & 7.9 & 161 & 7.7 & 117 & 7.7 & 147 & 8.5 & 127 & 7.7 \\ 
        Student and working & 72 & 1.0 & 23 & 1.1 & 14 & 0.9 & 18 & 1.0 & 17 & 1.0 \\ 
        On leave from work & 329 & 4.7 & 103 & 4.9 & 98 & 6.5 & 72 & 4.2 & 56 & 3.4 \\ 
        Seeking job & 395 & 5.6 & 119 & 5.7 & 93 & 6.1 & 87 & 5.0 & 96 & 5.8 \\ 
        Housekeeping & 1235 & 17.6 & 362 & 17.2 & 263 & 17.4 & 323 & 18.7 & 287 & 17.3 \\ 
        Student & 141 & 2.0 & 47 & 2.2 & 35 & 2.3 & 36 & 2.1 & 23 & 1.4 \\ 
        Other & 465 & 6.6 & 160 & 7.6 & 88 & 5.8 & 100 & 5.8 & 117 & 7.1 \\ 
        \hline
    \end{tabular}
    \caption{Demographic Attributes of the Respondents}
    \label{tab:Demographic}
    
\end{table}



\end{document}